\documentclass{article}
\usepackage{CJKutf8}
\usepackage[unicode]{hyperref} 
\usepackage{subcaption}

\usepackage{amsfonts}
\usepackage{amsmath}
\usepackage{amssymb}
\usepackage{amsthm}
\usepackage{youngtab}
\usepackage{braket}
\usepackage{graphicx}
\usepackage[svgnames]{xcolor}
\usepackage{ytableau,tikz}
\usepackage{here}
\usepackage{comment}
\usepackage{simplewick}
\usepackage{tikz}
\usepackage{feynmf}
\usepackage{mathtools}
\usepackage{qcircuit}

\newcommand{\ba}{\begin{eqnarray}}
\newcommand{\ea}{\end{eqnarray}}

 \makeatletter
\@addtoreset{equation}{section}

\makeatother

\title{A short review on TBA equation and scattering amplitude/Wilson loop duality\\
~\\
热力学Bethe拟设和散射振幅
}
\author{
 束红非 (Hongfei Shu)\footnote{shu@zzu.edu.cn}\footnote{国家自然科学基金(编号:12405087），河南省博士后经费（编号:2212005）和郑州大学启动经费（编号:21-35220581）资助项目。}\\
郑州大学物理学院天文所，河南省郑州市科学大道100号\\
Institute for Astrophysics, School of Physics, Zhengzhou University, Zhengzhou, Henan\\
450001, China 
}
\date{} 

\begin{document}

\begin{CJK*}{UTF8}{gbsn}

\maketitle

\begin{center}
{\bf 摘要/Abstract}
\end{center}

In this review, we introduce the computation of the minimal surface area in the scattering amplitude/Wilson loop duality, where the minimal surface ends on a light-like polygonal Wilson loop at the boundary of anti-de Sitter space (AdS). Due to its nonlinearity and the complexity of the boundary conditions, directly solving the equations of motion to compute the area is highly challenging. This paper reviews an alternative approach that bypasses the direct solution of the equations of motion and instead uses integrable systems to compute the area. We will provide boundary conditions for the Hitchin system, which is equivalent to the equations of motion, to describe the light-like polygonal boundary of the minimal surface. Starting from the solution of the Hitchin system, we will further derive the 
Y-system and the Thermodynamic Bethe Ansatz (TBA) equations, whose free energy provides the nontrivial part of the minimal surface area. Finally, we will discuss recent developments in this field and provide an outlook for future research.

~\\

本文旨在介绍散射振幅/Wilson圈对偶中极小曲面面积的计算，该极小曲面以反德西特空间（AdS）边界上的类光多边形Wilson圈为边界。鉴于其非线性以及复杂的边界条件，很难通过直接求解运动方程来计算此面积。文章将回顾如何绕开对运动方程的直接求解，转而利用可积系统来计算面积的方法。具体而言，将介绍与运动方程等价的Hitchin系统的边界条件，以描述极小曲面的类光多边形的边界。由此Hitchin系统的解出发，我们将进一步的推导出Y系统以及热力学 Bethe 拟设方程，其自由能将给出极小曲面面积的非平凡部分。最后，本文还将简要阐述该领域的后续发展以及展望。\\
~\\
关键词：可积系统，规范/弦对偶，非微扰方法

\newpage

\noindent\hrulefill
\tableofcontents
\noindent\hrulefill

\section{引言：散射振幅/Wilson圈对偶}

AdS/CFT-对偶\cite{Maldacena:1997re,Gubser:1998bc,Witten:1998qj}的发现为我们研究强耦合的规范场论提供了强有力的工具。得益于AdS$_5\times$ S$^5$中经典可积性以及平面极限下量子可积性的发现，四维最大超对称超杨-米尔斯理论（${\cal N}=4$ SYM）中谱、散射振幅和威尔逊圈（Wilson loop）等方面的研究取得了重大进展\cite{Bena:2003wd,Minahan:2002ve,Beisert:2010jr}。在这些进展当中，平面极限下的散射振幅/Wilson圈对偶的发现尤为引人注目。

\begin{figure}[hpt]
\centering
 \includegraphics[width=105mm]{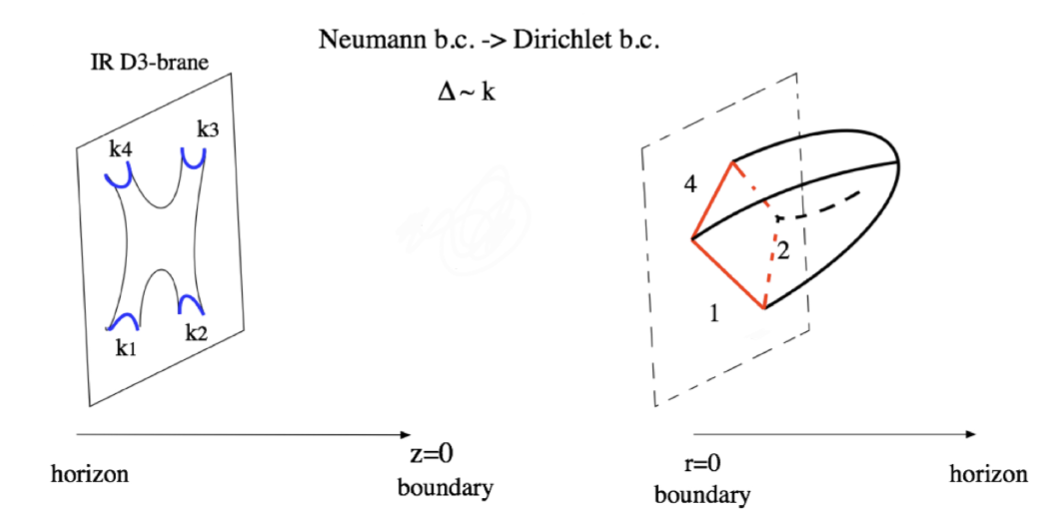}
    \caption{散射振幅与Wilson圈的对偶关系。左侧的D3膜延伸在空间$y^1,y^2,y^3$方向，并位于视界附近。在T-对偶后，我们得到了以类光多边形的Wilson圈为边界的世界面\\
    Figure 1: Scattering amplitude/Wilson loop
duality. The D3-brane on the left extends along the spatial directions $y^1, y^2, y^3$ and is located near the horizon. After T-duality, we obtain a worldsheet bounded by a light-like polygonal Wilson loop.}
    \label{fig:SA-WL}
\end{figure}

${\cal N}=4$ SYM中胶子的最大螺旋度破坏（Maximally helicity violating）散射可以由AdS空间中靠近视界的D3膜上的开弦散射来描述\footnote{这是因为我们关心散射振幅的IR发散，这个D3膜也被称作IR D3膜。}，如图\ref{fig:SA-WL}左侧所示。从这些开弦的世界面便能计算散射振幅。然而，由于这个世界面处于视界附近并且具有Neumann边界条件，直接计算较为困难。这里我们采用了AdS$_5$空间的Poincare坐标，即$d\tilde{s}^2=\frac{dz^2+d\vec{\tilde{y}}^2}{z^2}$。为了克服这一难题，Alday和Maldacena在$y^i$ ($i=0,\cdots3$)方向分别做了T-对偶\cite{Alday:2007hr}。引入新的径向坐标$r=1/z$后，T-对偶后的度规重新变回了AdS$_5$空间的度规，$ds^2=\frac{dr^2+d\vec{x}^2}{r^2}$。\footnote{更严格的说，此时的胀子(dilaton)以及5-形式会发生变化，为了抵消这些变化，我们需要在超空间做费米T-对偶\cite{Berkovits:2008ic}。}值得注意的是，由于径向方向的坐标变换，在新的度规下视界和边界的位置发生了互换。另外，在T-对偶下，原本世界面的Neumann边界条件变作了Dirichlet边界条件，并且动量变作了``缠绕数''，即右图红色线段的长度 $\Delta x^\mu=2\pi k^\mu$。由于胶子是无质量的以及散射过程中的动量守恒，$\Delta x^\mu$将给出一个类光的闭合多边形。因此，在（T-对偶后）AdS空间中，我们得到了一个以AdS边界上的类光多边形为边界的世界面，如图\ref{fig:SA-WL}右侧所示。根据AdS/CFT对应\cite{Rey:1998ik,Maldacena:1998im}，这个多边形对应类光多边形的Wilson圈，平面极限下强耦合的胶子散射振幅可以通过以类光多边形的Wilson圈为边界的极小曲面的面积来得到。这便是散射振幅/Wilson圈对偶。
这一发现不仅为强耦合的散射振幅的计算提供了新的思路，也揭示了平面极限下散射振幅的对偶超共形对称性的存在\cite{Beisert:2008iq}。

Alday和Maldacena通过求解运动方程，成功计算了以类光四边形Wilson圈为边界的极小曲面面积\cite{Alday:2007hr}，其对应于强耦合的四点胶子散射振幅，并且他们的计算结果验证了Bern-Dixon-Smirnov对胶子散射振幅IR发散的假说（BDS假说）\cite{Bern:2005iz} \footnote{由于篇幅限制，我们在此不做过多解释，请参见\cite{Alday:2008yw}。}。由于反常共形Ward恒等式的限制，四点及五点胶子散射振幅将满足BDS假说。高点散射振幅($n\geq 6$)将会与BDS假说存在偏离\cite{Drummond:2007au,Drummond:2008vq}。然而，由于世界面运动方程的非线性以及其极为复杂的边界条件，对于高点散射振幅的极小面积的计算变得十分困难\footnote{在\cite{Astefanesei:2007bk,Itoyama:2007ue,Itoyama:2007fs,Dobashi:2009sj},对于高点散射振幅的极小面积做了进一步的尝试，并发现了与BDS假设的偏离。}。

受到${\cal N}=2$越墙(wall-crossing)效应\cite{Gaiotto:2010okc,Gaiotto:2009hg}等相关研究的启发，Alday等人开创性地提出了一种无需直接求解运动方程即可计算极小曲面面积的方法\cite{Alday:2009yn,Alday:2009dv,Alday:2010vh,Hatsuda:2010cc}，这也是本文将探讨的主要内容。首先，我们利用Pohlmeyer约化\cite{Pohlmeyer:1975nb,DeVega:1992xc}，
将描述极小曲面的运动方程和Virasoro约束改写为特定的经典可积偏微分方程，其与Hitchin系统及其线性问题（linear problem）等价。通过设定这些线性问题的边界条件，我们能够重现出极小曲面的类光多边形Wilson圈边界。进一步的，通过组合线性问题的解，我们引入Y函数，该函数满足相应的热力学 Bethe 拟设（TBA）方程\footnote{这背后隐藏着常微分方程/量子可积系统（ODE/IM）对应的扩展\cite{Dorey:1998pt,Bazhanov:1998wj,Dorey:2007zx}。}。我们可以发现极小曲面的面积中的非平凡部分由TBA方程的自由能所决定。

本文的构造如下：在第\ref{sec:min-sur}节，我们首先介绍AdS$_3$空间经典弦的Pohlmeyer 约化，并将推导出修改sinh-Gordon方程以及相应的线性问题。我们将给出线性问题的边界条件，以重现极小曲面的类光多边形Wilson圈边界。在第\ref{sec:TBA}节中，我们给出相应的线性问题的小解，并以此为基础引入Y函数并推导出其所满足的TBA方程。在第\ref{sec:free-energy}节，通过TBA方程的自由能，我们将给出极小曲面面积的非平凡部分。最后，在第\ref{sec:con}节，我们将给出本文的总结以及展望。




\section{AdS$_3$空间中的极小曲面及其边界条件}\label{sec:min-sur}
在本章中，我们考虑AdS$_3$空间内的极小曲面，该极小曲面以AdS边界上的类光多边形Wilson圈为边界，见图\ref{fig:SA-WL}。我们需要求解经典的运动方程和Virasoro约束，以得到极小曲面的面积。然而，由于其非线性以及复杂边界条件，对于一般情形下的类光多边形Wilson圈上的世界面，直接求解其运动方程是非常困难的。

在本章中，我们从可积性的角度出发，首先考虑AdS空间中经典弦的Pohlmeyer约化，并得到与AdS$_3$中经典弦的运动方程与Viraoso约束等价的修改sinh-Gordon（MshG）方程。之后，我们考虑与MshG方程等价的Hitchin系统以及其线性问题，并给出其边界条件以刻画极小曲面的类光多边形Wilson圈边界。

\subsection{AdS空间中经典弦的Pohlmeyer约化}
经典弦的运动由其运动方程以及约束条件所决定。Pohlmeyer约化为解析这些复杂的方程提供了一种有效的途径\cite{Pohlmeyer:1975nb,DeVega:1992xc,Alday:2009yn,Papathanasiou:2012hx}。本章节，我们将介绍AdS时空中经典弦的Pohlmeyer约化。为了简化，我们主要介绍AdS$_3$的情形。

AdS$_d$空间可以被视作 $R^{(2,d-1)}$空间中的一个双曲面嵌入
\begin{equation}
    \vec{Y}\cdot\vec{Y}=-Y_{-1}^{2}-Y_{0}^{2}+Y_{1}^{2}+\cdots+Y_{d-1}^{2}=-1,
\end{equation}
其中$\vec{Y}$是$R^{(2,d-1)}$空间中的向量。弦的世界面的作用量为
\begin{equation}
    S=\int dz^{2}\Big(\partial\vec{Y}\cdot\bar{\partial}\vec{Y}-\Xi(z,\bar{z})(\vec{Y}\cdot\vec{Y}+1)\Big).
\end{equation}
这里 $z,\bar{z}$ 是世界面的坐标， $\Xi(z,\bar{z})$是拉格朗日乘子。从作用量，我们很容易得到世界面的运动方程以及Virasoro限制条件：
\begin{equation}\label{eq:eom}
    \partial\bar{\partial}\vec{Y}-(\partial\vec{Y}\cdot\bar{\partial}\vec{Y})\vec{Y}=0,\quad\partial\vec{Y}\cdot\partial\vec{Y}=0,\quad\bar{\partial}\vec{Y}\cdot\bar{\partial}\vec{Y}=0.
\end{equation}

我们首先引入以下AdS$_3$中$SO(2,2)$转动下不变的物理量：
\begin{equation}
    \begin{aligned}
        e^{2\alpha(z,\bar{z})}=&\frac{1}{2}\partial\vec{Y}\cdot\bar{\partial}\vec{Y},\quad N_{a}=\frac{e^{-2\alpha}}{2}\epsilon_{abcd}Y^{b}\partial Y^{c}\bar{\partial}Y^{d}\\p=&-\frac{1}{2}\vec{N}\cdot\partial^{2}\vec{Y},\quad p=-\frac{1}{2}\vec{N}\cdot\bar{\partial}^{2}\vec{Y}，\quad a=-1,0,1,2.
    \end{aligned}
\end{equation}
这些量满足 $\vec{N}\cdot\vec{N}=1, \vec{N}\cdot\vec{Y}=0,\vec{N}\cdot\partial\vec{Y}=0$ 以及 $\vec{N}\cdot\bar{\partial}\vec{Y}=0$。此外，通过\eqref{eq:eom}, 我们很容易发现$p=p(z)$是一个全纯函数。利用这些$SO(2,2)$不变量，我们可以引入一组正交基$(q_1,q_2,q_3,q_4)$:
\begin{equation}
    q_{1}=\vec{Y},\quad q_{2}=\frac{e^{-\alpha}}{2}(\bar{\partial}\vec{Y}-\partial\vec{Y}),\quad q_{3}=\frac{e^{-\alpha}}{2}(\bar{\partial}\vec{Y}+\partial\vec{Y}),\quad q_{4}=\vec{N},
\end{equation}
其满足 $q_i\cdot q_j={\rm diag}(-1,-1,1,1)$。利用运动方程及Virasoro约束\eqref{eq:eom}，我们可以计算这组正交基的微分
\begin{equation}
    \partial q_i=:B_{ij}q_j,\quad \bar{\partial} q_i=:A_{ij}q_j,
\end{equation}
其中，$A$和$B$是由$SO(2,2)$不变量写作的矩阵
\begin{equation}\label{eq:A-B}
   A=\left(\begin{array}{cccc}
0 & e^{\alpha} & e^{\alpha} & 0\\
-e^{\alpha} & 0 & \bar{\partial}\alpha & e^{-\alpha}\bar{p}\\
e^{\alpha} & \bar{\partial}\alpha & 0 & e^{-\alpha}\bar{p}\\
0 & e^{-\alpha}\bar{p} & -e^{-\alpha}\bar{p} & 0
\end{array}\right),\quad B=\left(\begin{array}{cccc}
0 & -e^{\alpha} & e^{\alpha} & 0\\
e^{\alpha} & 0 & -\partial\alpha & e^{-\alpha}p\\
e^{\alpha} & -\partial\alpha & 0 & -e^{-\alpha}p\\
0 & e^{-\alpha}p & e^{-\alpha}p & 0
\end{array}\right).
\end{equation}
从由自洽性条件（或可积条件） $\partial\bar{\partial}q_i=\bar{\partial}\partial q_i$，我们得到$A,B$的平坦性条件
\begin{equation}
    \partial A-\bar{\partial}B+[A,B]=0.
\end{equation}
将\eqref{eq:A-B}带入，我们得到与\eqref{eq:eom}等价的方程
\begin{equation}\label{eq:mshG}
    \begin{aligned}
        \partial\bar{\partial}\alpha-e^{2\alpha}+e^{-2\alpha}p\bar{p}=0,\quad \bar{\partial}p=\partial\bar{p}=0.
    \end{aligned}
\end{equation}
第一个条件被称作修改sinh-Gordon方程。

将此问题用旋量基重新表述是很有益处的。$SO(2,2)$群可以表达为两个$SL(2)$群的直积，即$SL(2)\times SL(2)$。时空向量$\vec{Y}$与内禀坐标$q_i$均能通过$SL(2)\times SL(2)$的双旋量来描述
\begin{equation}\label{eq:W-spinor}
    W_{\alpha\dot{\alpha},a\dot{a}}=\frac{1}{2}(q_{i})_{a\dot{a}}\tilde{\sigma}_{\alpha\dot{\alpha}}^{i}=\frac{1}{2}\left(\begin{array}{cc}
q_{1}-q_{4} & q_{3}+q_{2}\\
q_{3}-q_{2} & q_{1}+q_{4}
\end{array}\right)_{\alpha\dot{\alpha}},
\end{equation}
在这里我们分别用$a,\dot{a}$和$\alpha,\dot{\alpha}$标记内禀$SL(2)\times SL(2)$的指标与时空$SL(2)\times SL(2)$的指标。类似地，我们可以将运动方程写作
\begin{equation}\label{eq:spinor-partial}
    \begin{aligned}
        &\partial W_{\alpha\dot{\alpha},a\dot{a}}+{(B_{z}^{L})_{\alpha}}^{\beta}W_{\beta\dot{\alpha},a\dot{a}}+{(B_{z}^{R})_{\dot{\alpha}}}^{\dot{\beta}}W_{\alpha\dot{\beta},a\dot{a}}=0,\\
        &\bar{\partial}W_{\alpha\dot{\alpha},a\dot{a}}+{(B_{\bar{z}}^{L})_{\alpha}}^{\beta}W_{\beta\dot{\alpha},a\dot{a}}+{(B_{\bar{z}}^{R})_{\dot{\alpha}}}^{\dot{\beta}}W_{\alpha\dot{\beta},a\dot{a}}=0,
    \end{aligned}
\end{equation}
其中$B^{L,R}$是$SL(2)$的左右联络
\begin{equation}
    \begin{aligned}
        B_{z}^{L}=&\left(\begin{array}{cc}
\frac{1}{2}\partial\alpha & -e^{\alpha}\\
-e^{-\alpha}p(z) & -\frac{1}{2}\partial\alpha
\end{array}\right),\quad B_{\bar{z}}^{L}=\left(\begin{array}{cc}
-\frac{1}{2}\bar{\partial}\alpha & -e^{-\alpha}\bar{p}(\bar{z})\\
-e^{\alpha} & \frac{1}{2}\bar{\partial}\alpha
\end{array}\right),\\
B_{z}^{R}=&\left(\begin{array}{cc}
-\frac{1}{2}\partial\alpha & -e^{-\alpha}p(z)\\
-e^{\alpha} & \frac{1}{2}\partial\alpha
\end{array}\right),\quad B_{\bar{z}}^{R}=\left(\begin{array}{cc}
\frac{1}{2}\bar{\partial}\alpha & -e^{\alpha}\\
e^{-\alpha}\bar{p}(\bar{z}) & -\frac{1}{2}\bar{\partial}\alpha
\end{array}\right).
    \end{aligned}
\end{equation}
由\eqref{eq:spinor-partial}的自洽性条件，我们发现联络$B$是平坦的，即
\begin{equation}
    \partial B_{\bar{z}}^{L,R}-\bar{\partial}B_{z}^{L,R}+[B_{z}^{L,R},B_{\bar{z}}^{L,R}]=0.
\end{equation}
同样的，平坦条件给出了修改sinh-Gordon方程\eqref{eq:mshG}。

\subsubsection{经典弦的重构}
我们可以通过修改sinh-Gordon方程\eqref{eq:mshG}的解得到$q_i$，并据此构建出经典弦的解。首先，我们利用$B^{L,R}$引入辅助线性问题(linear problem)
\begin{equation}
    \begin{aligned}
        &\partial\psi_{\alpha}^{L}+{(B_{z}^{L})_{\alpha}}^{\beta}\psi_{\beta}^{L}=0,\quad\bar{\partial}\psi_{\alpha}^{L}+{(B_{\bar{z}}^{L})_{\alpha}}^{\beta}\psi_{\beta}^{L}=0,\\
        &\partial\psi_{\dot{\alpha}}^{R}+{(B_{z}^{R})_{\dot{\alpha}}}^{\dot{\beta}}\psi_{\dot{\beta}}^{R}=0,\quad\bar{\partial}\psi_{\dot{\alpha}}^{R}+{(B_{\bar{z}}^{R})_{\dot{\alpha}}}^{\dot{\beta}}\psi_{\dot{\beta}}^{R}=0.
    \end{aligned}
\end{equation}
每组线性问题都会有两个独立解，记作$\psi^L_{\alpha,a}$和 $\psi^R_{\dot{\alpha},\dot{a}}$。张量\eqref{eq:W-spinor}的每一个成分都是零向量，并利用线性问题的解的唯一性，我们总能将\eqref{eq:W-spinor}表达为
\begin{equation}
    W_{\alpha\dot{\alpha},a\dot{a}}^{L}=\psi_{\alpha,a}^{L}\psi_{\dot{\alpha},\dot{a}}^{R}
\end{equation}
从$W_{\alpha\dot{\alpha},a\dot{a}}^{L}$，我们可以提取出$q_1$，并得到经典弦的坐标
\begin{equation}\label{eq:st-sol-linear}
    Y_{a\dot{a}}=\left(\begin{array}{cc}
Y_{-1}+Y_{2} & Y_{1}-Y_{0}\\
Y_{1}+Y_{0} & Y_{-1}-Y_{2}
\end{array}\right)_{a\dot{a}}=\psi_{\alpha,a}^{L}M_{1}^{\alpha\dot{\beta}}\psi_{\dot{\beta}\dot{a}}^{R},\quad M_{1}^{\alpha\dot{\beta}}=\left(\begin{array}{cc}
1 & 0\\
0 & 1
\end{array}\right).
\end{equation}

\subsubsection{Hitchin系统}

通过引入谱参数$\zeta\in\mathbb{C}$, 我们可以将左联络$B^L$扩展为一族平坦联络：
\begin{equation}
    B_{z}(\zeta)=\left(\begin{array}{cc}
\frac{1}{2}\partial\alpha & -\frac{1}{\zeta}e^{\alpha}\\
-\frac{1}{\zeta}e^{-\alpha}p(z) & -\frac{1}{2}\partial\alpha
\end{array}\right),\quad B_{\bar{z}}(\zeta)=\left(\begin{array}{cc}
-\frac{1}{2}\bar{\partial}\alpha & -\zeta e^{-\alpha}\bar{p}(\bar{z})\\
-\zeta e^{\alpha} & \frac{1}{2}\bar{\partial}\alpha
\end{array}\right).
\end{equation}
原始的左右联络$B^{L,R}$可以从$B_z(\zeta)$得到
\begin{equation}
    B_{z}^{L}=B_{z}(1),\quad B_{z}^{R}=UB_{z}(i)U^{-1},\quad U=\left(\begin{array}{cc}
0 & e^{i\pi/4}\\
e^{3i\pi/4} & 0
\end{array}\right).
\end{equation}
若将$B_z(\zeta),B_{\bar{z}}(\zeta)$分解为两部分，该问题的可积性将更显而易见
\begin{equation}
    B_z(\zeta)=A_z+\frac{1}{\zeta}\Phi_z,\quad B_{\bar{z}}(\zeta)=A_{\bar{z}}+\zeta \Phi_{\bar{z}},
\end{equation}
其中，$A,\Phi$是
\begin{equation}
    \begin{aligned}
        A_{z}=&\left(\begin{array}{cc}
\frac{1}{2}\partial\alpha & 0\\
0 & -\frac{1}{2}\partial\alpha
\end{array}\right),\quad\Phi_{z}=\left(\begin{array}{cc}
0 & -e^{\alpha}\\
-e^{-\alpha}p(z) & 0
\end{array}\right),\\
A_{\bar{z}}=&\left(\begin{array}{cc}
-\frac{1}{2}\bar{\partial}\alpha & 0\\
0 & \frac{1}{2}\bar{\partial}\alpha
\end{array}\right),\quad\Phi_{\bar{z}}=\left(\begin{array}{cc}
0 & -e^{-\alpha}\bar{p}(\bar{z})\\
-e^{\alpha} & 0
\end{array}\right).
    \end{aligned}
\end{equation}
由于平坦条件$\partial B_z-\bar{\partial}B_z+[B_z,B_{\bar{z}}]=0$对于所有的谱参数$\zeta$都成立，我们可以发现
\begin{equation}
    D_{z}\Phi_{\bar{z}}=D_{\bar{z}}\Phi_{z}=0,\quad F_{z\bar{z}}+[\Phi_{z},\Phi_{\bar{z}}]=0,
\end{equation}
其中
\begin{equation}
    \begin{aligned}
        D_{z}\Phi_{\bar{z}}=&\partial_{z}\Phi_{\bar{z}}+[A_{z,}\Phi_{\bar{z}}],\quad D_{\bar{z}}\Phi_{z}=\partial_{\bar{z}}\Phi_{z}+[A_{\bar{z},}\Phi_{z}],\\&F_{\bar{z}\bar{z}}=\partial A_{\bar{z}}-\bar{\partial}A_{z}+[A_{z},A_{\bar{z}}].
    \end{aligned}
\end{equation}
这对应于特定$SU(2)$规范群的Hitchin系统\footnote{Hitchin系统通过将四维自对偶条件（瞬子方程）降维到二维而产生的。$A_z$,$A_{\bar{z}}$和$\Phi_{z}$,$\Phi_{\bar{z}}$分别对应于二维的规范联络和Higgs场。}，并与与以下线性问题等价
\begin{equation}\label{eq:linear-pro}
    \partial_{z}\psi+B_{z}(\zeta)\psi=0, \quad \partial_{\bar{z}}\psi+B_{\bar{z}}(\zeta)\psi=0.
\end{equation}
方程\eqref{eq:st-sol-linear}可以被重写为
\begin{equation}
    \label{st-cd}
\left(\begin{array}{cc}
Y_{-1}+Y_{4} & Y_{1}-Y_{0}\\
Y_{1}+Y_{0} & Y_{-1}-Y_{4}
\end{array}\right)_{a,\dot{a}}=\psi_{\alpha,a}(\zeta=1)M_{\alpha\dot{\beta}}\psi_{\dot{\beta},\dot{a}}(\zeta=i),
\end{equation}
其中，矩阵$M$的具体形式取决于我们选取的左右线性问题。

\subsection{线性问题及世界面的边界条件}
为了深入探究多边形边界的极小曲面，利用线性问题将十分便捷。首先，我们聚焦于修改sinh-Gordon方程\eqref{eq:mshG}的一个特解$p=1,\alpha=0$。通过规范变换，我们可以将线性问题转化为
\begin{equation}
   (\partial_w-\zeta^{-1}\sigma_3)\hat{\psi}=0,\quad  ({\partial}_{\bar{w}}-\zeta\sigma_3)\hat{\psi}=0,
\end{equation}
其独立解可以写作$\hat{\eta}_+(\zeta)=(e^{\frac{w}{\zeta}+\bar{w}\zeta},0)^T, \hat{\eta}_-(\zeta)=(0,e^{\frac{w}{\zeta}+\bar{w}\zeta})^T$。 
$\hat{\eta}_+^L=\hat{\eta}_+(\zeta=1)$与$\hat{\eta}_-^L$分别在${\rm Re}(w)\to \infty$与${\rm Re}(w)\to -\infty$处指数增长。此时复平面上有两个（anti-Stokes）分段（sector），它们以${\rm Re}(w)=0$为分界。相应地，右线性问题有以${\rm Im}(w)=0$为边界的两个分段。从\eqref{st-cd}可以观察到，经典弦的世界面共有四个分段构成，这恰好描述了四边形边界条件。

进一步，我们考虑更复杂的$p(z)=z^{n-2}+\cdots$，为了简化我们引入新坐标$dw=\sqrt{p(z)}dz,d\bar{w}=\sqrt{\bar{p}(\bar{z})}d\bar{z}$，\eqref{eq:mshG}将变作
\begin{equation}
    \partial_{w}\bar{\partial}_{\bar{w}}\hat{\alpha}(w,\bar{w})-e^{2\hat{\alpha}(w,\bar{w})}+e^{-2\hat{\alpha}(w,\bar{w})}=0,
\end{equation}
其中$\hat{\alpha}=\alpha-\frac{1}{4}\log p\bar{p}$。当$\hat{\alpha}\to 0， z\to \infty$时，我们便在$w-$坐标再现之前特解的情形，每一页上将会有四个分段出现。在$z-$坐标，大$z$极限下我们可以看到有$2n$个分段，这描述了$2n$边形的边界条件。

综上所述，与散射振幅/类光多边形威尔逊环相关的极小曲面由一个多项式$p(z)$及$\alpha$的边界条件所决定
\begin{equation}
  p(z)=z^{n-2}+\cdots,\quad   e^{2\alpha}\sim \sqrt{p\bar{p}},\quad z\to \infty.
\end{equation}
为了深入刻画边界条件，我们考虑线性问题的解
\begin{equation}
    \psi_a=c_a^{\rm b}b+c_a^{\rm s}s,
\end{equation}
其总能表示为指数增大的``大解''$b$和指数压低的``小解''$s$的线性组合，$c_a^{\rm b,s}$是相应的系数。由于我们总能将小解的一部分加到大解中，因此系数$c_a^{\rm s}$并没有没有明确定义。相反的，一旦我们为小解选择了一个归一化标准，大解的系数$c_a^{\rm b}$就可以明确定义了。在此，我们将归一化条件设定为$b\wedge s=1$，$c_a^{\rm b}$便可以用$\psi_a$与$s$的内积得到
\begin{equation}
    \psi_a\wedge s=c_a^{\rm b}.
\end{equation}
在大$z$，经典弦的坐标\eqref{st-cd}可以近似为
\begin{equation}
    Y_{a\dot{a}}\sim(\psi_{a}^{L}\wedge s^{L})(\psi_{\dot{a}}^{R}\wedge s^{R})(b_{\alpha}^{L}M_{1}^{\alpha\dot{\beta}}b_{\dot{\beta}}^{R})\sim \lambda_a\tilde{\lambda}_{\dot{a}},
\end{equation}
其中$\lambda_a\sim \psi_a^L\wedge s^L, \tilde{\lambda}_{\dot{a}}\sim \psi_{\dot{a}}^R\wedge s^R$。由此可得$\vec{Y}$的内积为
\begin{equation}
    \vec{Y}^{2}=-\frac{1}{2}Y_{a\dot{a}}Y_{b\dot{b}}\epsilon^{ab}\epsilon^{\dot{a}\dot{b}}\sim\frac{1}{2}\lambda_{a}\tilde{\lambda}_{\dot{a}}\lambda_{b}\tilde{\lambda}_{\dot{b}}\epsilon^{ab}\epsilon^{\dot{a}\dot{b}}=0,
\end{equation}
这意味着大$z$极限下，经典弦在AdS时空的共形边界$\vec{Y}^2=0$处。引入Poincare坐标
\begin{equation}
    Y_{-1}+Y_2=\frac{1}{r},\quad x^\pm =\frac{Y_1\pm Y_0}{Y_{-1}+Y_2},
\end{equation}
可进一步发现
\begin{equation}
    x_{ij}^{+}:=x_{i}^{+}-x_{j}^{+}=\frac{\psi_{2}^{L}\wedge s_{i}^{L}}{\psi_{1}^{L}\wedge s_{i}^{L}}-\frac{\psi_{2}^{L}\wedge s_{j}^{L}}{\psi_{1}^{L}\wedge s_{j}^{L}}=-\frac{s_{i}^{L}\wedge s_{j}^{L}}{(\psi_{1}^{L}\wedge s_{i}^{L})(\psi_{1}^{L}\wedge s_{j}^{L})},
\end{equation}
其中$i,j$是多边形顶点的标记。由此我们得到交比(cross-ratio)\footnote{这里的交比是由Wilson圈的坐标写出的。通过T-对偶前后坐标空间与动量空间的关系，即$x_{i+1}-x_i=p_i$，我们可以将这些交比翻译为散射振幅的动量形式的交比。}
\begin{equation}\label{eq:cross-ratio-1}
    \frac{x_{ij}^{+}x_{kl}^{+}}{x_{ik}^{+}x_{jl}^{+}}=\frac{s_{i}^{L}\wedge s_{j}^{L}s_{k}^{L}\wedge s_{l}^{L}}{s_{i}^{L}\wedge s_{k}^{L}s_{j}^{L}\wedge s_{l}^{L}}=:\chi_{ijkl}(\zeta=1)
\end{equation}
同理，可得$x^-$方向的交比为
\begin{equation}\label{eq:cross-ratio-2}
    \frac{x_{ij}^{-}x_{kl}^{-}}{x_{ik}^{-}x_{jl}^{-}}=\frac{s_{i}^{R}\wedge s_{j}^{R}s_{k}^{R}\wedge s_{l}^{R}}{s_{i}^{R}\wedge s_{k}^{R}s_{j}^{R}\wedge s_{l}^{R}}=:\chi_{ijkl}(\zeta=i).
\end{equation}
当$p\sim z^{n-2}$时，左右线性问题分别有$n$个（anti-Stokes）分段。此多边形的顶点可以被标记为$(x_i^+,x_{i-1}^-)$, $(x_i^+,x_{i}^-)$, $(x_{i+1}^+,x_{i}^-)$, $ \cdots$。由此我们得到了在AdS边界处的类光$2n$多边形。从交比，我们可以进一步读取光多边形的形状。左右交比的总个数为$2(n-3)$，这正好与多项式$p(z)$的自由度一致\footnote{通过$z$的平移，我们总能将$z^{n-3}$的系数设为0。利用缩放，可将$z^{n-2}$的系数设为1。这样多项式中会有$n-3$个复参数出现。}。另外，$2(n-3)$也是2维空间中$2n$边形Wilson圈的自由度的数目，这里$2\times3$代表$SO(2,2)$的自由度。散射振幅/极小面积等物理量最终都应通过交比来表示。

本章节的讨论可以很容易的扩展到形状因子$\braket{k^{\rm in}_i|{\cal O}(q)|k^{\rm out}_j}$的情形。由于算符${\cal O}(q)$携带的动量，胶子的总动量$q\neq 0$，T-对偶后我们将得到一个周期性的Wilson线（Wilson line）\cite{Alday:2007he}。另外，在算符插入点是界面将不再是顺滑的，相应的我们需要在$p(z)$中引入奇异点\cite{Maldacena:2010kp,Gao:2013dza}。通过，修改Hitchin系统的边界条件，便能重现相应的极小曲面。

\section{热力学 Bethe 拟设方程与极小曲面的面积}\label{sec:TBA}
在上述章节中，我们认识到小解$s$在刻画物理量中的重要性，本章节中将深入探讨线性问题的Stokes现象以及相关的小解。这些小解的组合将满足T-/Y系统以及相应的热力学 Bethe 拟设方程，这些方程决定了极小曲面面积的非平凡部分。

\subsection{线性问题的小解}
我们考虑具有以下边界条件的线性问题
\begin{equation}\label{eq:bdYcond}
     e^{2\alpha}\sim \sqrt{p\bar{p}},\quad z\to \infty,\quad p(z)=z^{n-2}+c_{n-4}z^{n-4}+\cdots+c_{0}.
\end{equation}
对于任意谱参量$\zeta$，$w$平面的(anti-Stokes)分段是\footnote{这可以从上述$\hat{\eta}_\pm(\zeta)$读出。}
\begin{equation}
    \hat{W}_{j}[\zeta]:(j-\frac{3}{2})\pi+{\rm arg}(\zeta)<{\rm arg}(w)<(j-\frac{1}{2})\pi+{\rm arg}(\zeta).
\end{equation}
在$z$平面，这些分段可以写作
\begin{equation}
    W_{j}[\zeta]:(j-\frac{3}{2})\frac{2\pi}{n}+\frac{2}{n}{\rm arg}(\zeta)<{\rm arg}(z)<(j-\frac{1}{2})\frac{2\pi}{n}+\frac{2}{n}{\rm arg}(\zeta).
\end{equation}
容易发现联络$B(\zeta)$满足以下$\mathbb{Z}_2$变换：
\begin{equation}
    \sigma_3B_{z,\bar{z}}(\zeta)\sigma_3=B_{z,\bar{z}}(e^{i\pi}\zeta),
\end{equation}
线性问题从而变换为
\begin{equation}
    (\partial_z+B_z)\sigma^3\psi(z,\bar{z}|e^{i\pi}\zeta)=0.
\end{equation}
从分段$W_1$的小解$s_1(\zeta)$，我们可以生成新的解
\begin{equation}
    s_{k+1}=(i\sigma^3)^ks_1(e^{i\pi k}\zeta),
\end{equation}
通过这些解的渐近行为，可以发现它们分别是分段$W_{k+1}$上的小解。这些小解的$SL(2)$不变的内积独立于坐标$z$但依存于$\zeta$，并满足以下性质\footnote{此内积可以理解为两个向量组成的行列的行列式。}
\begin{equation}\label{eq:s-prod-prop}
    (s_{i}\wedge s_{k})(e^{i\pi} \zeta)=(s_{i+1}\wedge s_{k+1})( \zeta),\quad  s_{i}\wedge s_{i+1}=1.
\end{equation}
此外，这些内积满足行列式的Schouten恒等式：
\begin{equation}\label{eq:s-iden}
    s_{i}\wedge s_{j}s_{k}\wedge s_{l}+s_{i}\wedge s_{l}s_{j}\wedge s_{k}+s_{i}\wedge s_{k}s_{l}\wedge s_{j}=0.
\end{equation}

\subsection{Fock–Goncharov坐标及Y系统}
\begin{figure}[hpt]
\includegraphics[width=\linewidth]{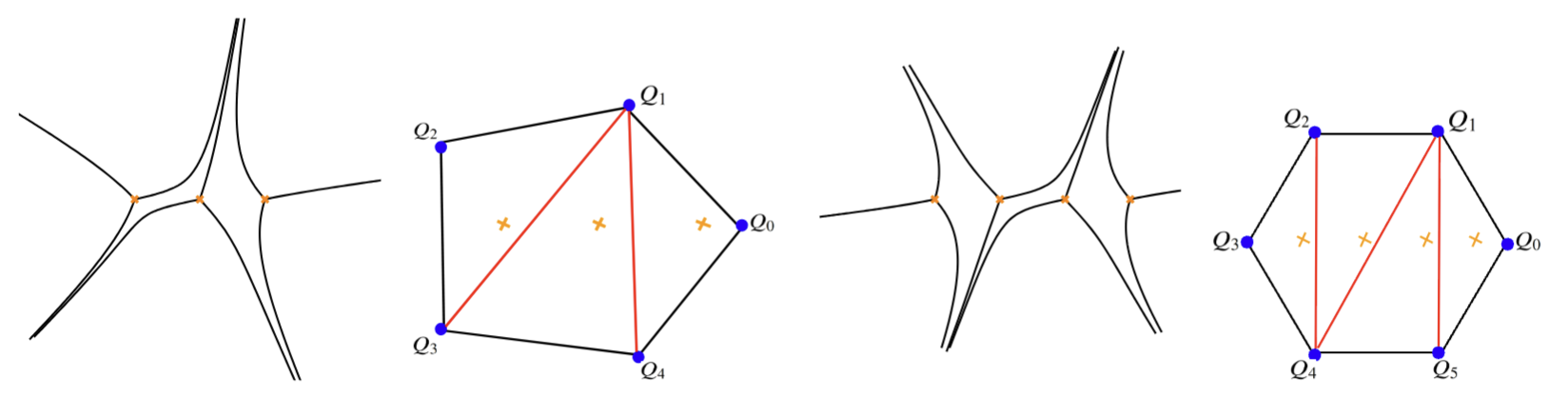}
       \caption{$n=5$ 及$n=6$ 的WKB曲线以及对应的WKB三角分割。这里叉号代表$p(z)$的零点。简单起见，我们令所有的零点为实数。另外，我们对$\zeta$做了微小的转动，以避免连结零点和零点的WKB曲线。\\
       Figure 2: The WKB curves and corresponding WKB triangulations for $n=5$ and $n=6$ . Here, the cross marks represent the zeros of $p(z)$. For simplicity, we assume all zeros to be real. Additionally, a slight rotation of $\zeta$ is applied to avoid WKB curves connecting zeros to zeros.}
    \label{fig:Stokes}
\end{figure}

通过一些规范变换，总能将线性问题的全纯部分变为
\begin{equation}
   \Big( \partial_z+{\rm diag}(\sqrt{p}, -\sqrt{p})\Big)\tilde{\psi}=0.
\end{equation}
利用WKB近似，我们得到$s_k$的渐近行为
$s_k\sim C\exp\big(\frac{\delta_k}{\zeta}\int_{z_{k}}^{z}\sqrt{p}dz\big)$。我们可以很自然的引入黎曼面$y^2=p(z)$。为保证WKB近似的准确性，我需要利用由以下参量$t$刻画的WKB曲线上的解
\begin{equation}
    {\rm Im}(\frac{1}{\zeta}\sqrt{p(z)}\frac{dz}{dt})=0.
\end{equation}
在图\ref{fig:Stokes}的（第1分图）和(第3分图）中，我们分别画出$n=5$及$n=6$的WKB曲线。在复$z$-平面上一般点处，WKB曲线并不会相交。从$p(z)$的简单零点出发会辐射出三条WKB曲线，其将把复平面分割为三个区域。另外，$z=\infty$是一个$n+2$的非正规奇异点(irregular singular point)，记作$Q$。$Q$周围的WKB曲线将把平面分割为$n$个区域。我们可以很方便的将$Q$理解为$n$个被标记的奇异点，$Q_k$, $k=1,\cdots,n$。为保证其唯一性，我们将$Q_k$放在小解$s_k$衰减的最快的位置。
通过以上操作，我们会发现复平面被WKB曲线分割为数个单元格(cell)。每个单元格中将会有无数条同伦等价的曲线连结两个(被标记的)奇异点。选取每个单元格的代表曲线，我们将得到所谓的WKB三角分割(WKB triangulation)，记作$T_{\rm WKB}$。与图\ref{fig:Stokes}（第1分图）和(第3分图）对应的WKB三角分割参见图\ref{fig:Stokes}的(第2分图)和(第4分图)。这些WKB三角的顶点位于正规奇异点或者非正规奇异点的被标记点处。在每一个顶点处将会有至少一条边(edge)。如果两个三角有共同的边$E$，其将组合成一个四边形(quadrilateral)，记作${\cal Q}_E$。

给定一个四边形${\cal Q}_E$，假设其顶点$Q_1,\cdots,Q_4$沿逆时针排列，并且两个三角形共同的边为$E(Q_1,Q_3)$。我们引入Fock–Goncharov坐标
\begin{equation}
    {\cal X}_{E(Q_{1},Q_{3})}=-\frac{(s_{1}\wedge s_{2})(s_{3}\wedge s_{4})}{(s_{2}\wedge s_{3})(s_{4}\wedge s_{1})}.
\end{equation}
基于此规则，我们可以写出图\ref{fig:Stokes} (b) 的两个非平凡的Fock–Goncharov坐标\cite{Gaiotto:2009hg}
\begin{equation}
    {\cal X}_{E(Q_{1},Q_{4})}^{(b)}=-\frac{(s_{1}\wedge s_{3})(s_{4}\wedge s_{0})}{(s_{3}\wedge s_{4})(s_{0}\wedge s_{1})},\quad {\cal X}_{E(Q_{1},Q_{3})}^{(b)}=-\frac{(s_{1}\wedge s_{2})(s_{3}\wedge s_{4})}{(s_{2}\wedge s_{3})(s_{4}\wedge s_{1})}.
\end{equation}
图.\ref{fig:Stokes} (d) 则有三个非平凡的Fock–Goncharov坐标
\begin{equation}
    {\cal X}_{E(Q_{1},Q_{5})}^{(d)}=-\frac{(s_{1}\wedge s_{4})(s_{5}\wedge s_{0})}{(s_{4}\wedge s_{5})(s_{0}\wedge s_{1})},\quad {\cal X}_{E(Q_{1},Q_{4})}^{(d)}=-\frac{(s_{1}\wedge s_{2})(s_{4}\wedge s_{5})}{(s_{2}\wedge s_{4})(s_{5}\wedge s_{1})},\quad {\cal X}_{E(Q_{2},Q_{4})}^{(d)}=-\frac{(s_{2}\wedge s_{3})(s_{4}\wedge s_{1})}{(s_{3}\wedge s_{4})(s_{1}\wedge s_{2})}.
\end{equation}
对于任意的$n$，这些Fock–Goncharov坐标可以系统的表达为\footnote{利用$s_{k}\propto s_{n+k}$，我们很容易发现所有的Fock–Goncharov坐标都可以有Y-函数表达，例如${\cal X}_{E(Q_{1},Q_{4})}^{(b)}=Y_{1}^{[-]}$, ${\cal X}_{E(Q_{1},Q_{3})}^{(b)}=Y_{2}^{-1}$。更为详细的联系可参见\cite{Ito:2025pfo}。
}
\begin{equation}\label{eq:Y}
    \begin{aligned}
        Y_{2k}=\frac{(s_{-k}\wedge s_{k})(s_{-k-1}\wedge s_{k+1})}{(s_{-k-1}\wedge s_{-k})(s_{k}\wedge s_{k+1})},\quad Y_{2k+1}^{[-]}=\frac{(s_{-k-1}\wedge s_{k})(s_{-k-2}\wedge s_{k+1})}{(s_{-k-2}\wedge s_{-k-1})(s_{k}\wedge s_{k+1})},
    \end{aligned}
\end{equation}
这里$f^{[k]}=f(e^{k \frac{\pi i}{2}}\zeta)$。
利用\eqref{eq:s-prod-prop}以及Schouten恒等式\eqref{eq:s-iden}，我们发现Y函数满足的Y系统\footnote{引入$T_{n-1}=(s_0\wedge s_n)^{[-n]}=0$，我们也很容易发现$T_n$满足可积系统里的Hirota方程。Y-函数可以表示为$Y_n=T_{n-1}T_{n+1}$。}
\begin{equation}
    Y_{k}^{[-]}Y_{k}^{[+]}=(1+Y_{k-1})(1+Y_{k+1}),\quad k=1,2,\cdots,n-3.
\end{equation}
这是量子可积系统中的$A_{n-3}$ Y系统\cite{Zamolodchikov:1991et}。让$\zeta=1,i$，我们可以从Y函数得到相应的交比\eqref{eq:cross-ratio-1} 和\eqref{eq:cross-ratio-2}。
值得强调的是Fock–Goncharov坐标的定义只依赖于WKB曲线，其定义也适用于形状因子或者非平面散射振幅等对应的情形。

\subsection{热力学 Bethe 拟设方程}
在量子可积系统中，给定Y函数的解析性后，可以将Y系统转化为热力学 Bethe 拟设（TBA）方程。在本节，我们将利用WKB分析计算Y函数在大$\zeta$和小$\zeta$的渐近行为，并推导TBA方程。

在小$\zeta$时，得到$s_k$的渐近行为是
\begin{equation}
    s_k\sim C\exp\big(\frac{\delta_k}{\zeta}\int_{z_{k}}^{z}\sqrt{p}dz\big),
\end{equation}
其中$C$是一个常数向量，$z_k$是积分起点。$\delta_k=\pm(-1)^k$中的正负号取决于小解所在的黎曼面。当积分的径路穿过黎曼面上的割线（branch cut）时，这个符号会变化。进一步地，我们可以分析$s_i\wedge s_j$的渐近行为
\begin{equation}
s_{i}\wedge s_{j}\sim\exp\Big(\frac{\delta_{i}}{\zeta}\int_{z_{i}}^{z_{j}}\sqrt{p}dz+\cdots\Big),
\end{equation}
其中黎曼面的选择，总能让$\delta_i=-\delta_j$。Y函数\eqref{eq:Y}包含4个内积，可以发现相应的4条积分路径总能组合为一个闭合的圈$\gamma$~\footnote{通过画出线性问题的Stokes图，我们可以更系统的分析Y函数的渐近行为。这里由于篇幅所限，我们不展开详细分析。感兴趣的读者可以参考\cite{Gaiotto:2009hg,Alday:2010vh}。}
\begin{equation}
    \label{eq:Yasym}
\begin{aligned}
\log{Y}_{2j+1}(\zeta,u_{a})&\sim\frac{1}{i\zeta}\oint_{\gamma_{2j+1}}p_0(x)dx=:-\frac{m_{2j+1}}{2\zeta},\\
\log{Y}_{2j}(\zeta,u_{a})&\sim\frac{1}{\zeta}\oint_{\gamma_{2j}}p_0(x)dx=:-\frac{m_{2j}}{2\zeta},\quad \zeta \to 0,
\end{aligned}
\end{equation}
其中圈$\gamma_i$见图\ref{fig:cycle}。同样地，我们可以得到大$\zeta$时的Y函数的渐近行为$\log Y_k(\zeta)\to \frac{m_k}{2}\zeta+\cdots$。综上所述，我们可以得到Y函数的渐近行为
\begin{equation}
    \log Y_k=-m_s\cosh\theta,\quad\theta\to \pm \infty,
\end{equation}
其中$\zeta=e^\theta$。

\begin{figure}[htb]
\begin{center}
    
\tikzset{every picture/.style={line width=0.75pt}} 

\begin{tikzpicture}[x=0.55pt,y=0.55pt,yscale=-1,xscale=1]

\draw    (154.45,102.32) -- (167.04,114.68) ;
\draw    (154.93,115.32) -- (166.08,103.52) ;
\draw    (244.45,102.32) -- (257.04,114.68) ;
\draw    (244.93,115.32) -- (256.08,103.52) ;
\draw    (344.45,102.32) -- (357.04,114.68) ;
\draw    (344.93,115.32) -- (356.08,103.52) ;
\draw    (434.45,102.32) -- (447.04,114.68) ;
\draw    (434.93,115.32) -- (446.08,103.52) ;
\draw  [color={rgb, 255:red, 0; green, 0; blue, 0 }  ,draw opacity=1 ][line width=1.5]  (135.17,113) .. controls (135.17,94.22) and (166.17,79) .. (204.42,79) .. controls (242.66,79) and (273.67,94.22) .. (273.67,113) .. controls (273.67,131.78) and (242.66,147) .. (204.42,147) .. controls (166.17,147) and (135.17,131.78) .. (135.17,113) -- cycle ;
\draw    (351.5,109) .. controls (353.43,107.36) and (355.3,107.39) .. (357.1,109.08) .. controls (358.82,110.77) and (360.52,110.8) .. (362.21,109.16) .. controls (363.82,107.51) and (365.38,107.53) .. (366.87,109.22) .. controls (368.3,110.91) and (369.93,110.93) .. (371.77,109.29) .. controls (373.5,107.64) and (375.15,107.66) .. (376.72,109.34) .. controls (378.55,111.03) and (380.16,111.04) .. (381.54,109.38) .. controls (383.43,107.73) and (385.21,107.74) .. (386.9,109.41) .. controls (388.43,111.08) and (389.98,111.08) .. (391.57,109.41) .. controls (393.35,107.72) and (395.04,107.7) .. (396.64,109.34) .. controls (398.27,110.97) and (399.94,110.92) .. (401.66,109.19) .. controls (403.22,107.46) and (404.91,107.39) .. (406.74,108.98) .. controls (408.43,110.57) and (410.05,110.5) .. (411.61,108.77) .. controls (413.16,107.04) and (414.87,106.98) .. (416.75,108.58) .. controls (418.54,110.19) and (420.21,110.14) .. (421.76,108.43) .. controls (423.46,106.72) and (425.09,106.67) .. (426.64,108.3) .. controls (428.29,109.93) and (429.96,109.89) .. (431.65,108.18) .. controls (433.46,106.47) and (435.11,106.44) .. (436.6,108.07) -- (440.5,108) ;
\draw    (161.5,110) .. controls (163.43,108.36) and (165.3,108.39) .. (167.1,110.08) .. controls (168.82,111.77) and (170.52,111.8) .. (172.21,110.16) .. controls (173.82,108.51) and (175.38,108.53) .. (176.87,110.22) .. controls (178.3,111.91) and (179.93,111.93) .. (181.77,110.29) .. controls (183.5,108.64) and (185.15,108.66) .. (186.72,110.34) .. controls (188.55,112.03) and (190.16,112.04) .. (191.54,110.38) .. controls (193.43,108.73) and (195.21,108.74) .. (196.9,110.41) .. controls (198.43,112.08) and (199.98,112.08) .. (201.57,110.41) .. controls (203.35,108.72) and (205.04,108.7) .. (206.64,110.34) .. controls (208.27,111.97) and (209.94,111.92) .. (211.66,110.19) .. controls (213.22,108.46) and (214.91,108.39) .. (216.74,109.98) .. controls (218.43,111.57) and (220.05,111.5) .. (221.61,109.77) .. controls (223.16,108.04) and (224.87,107.98) .. (226.75,109.58) .. controls (228.54,111.19) and (230.21,111.14) .. (231.76,109.43) .. controls (233.46,107.72) and (235.09,107.67) .. (236.64,109.3) .. controls (238.29,110.93) and (239.96,110.89) .. (241.65,109.18) .. controls (243.46,107.47) and (245.11,107.44) .. (246.6,109.07) -- (250.5,109) ;
\draw [color={rgb, 255:red, 0; green, 0; blue, 0 }  ,draw opacity=1 ][line width=1.5]    (236.5,109) .. controls (248.5,64) and (353.5,65) .. (368.5,109) ;
\draw [color={rgb, 255:red, 0; green, 0; blue, 0 }  ,draw opacity=1 ][line width=1.5]  [dash pattern={on 1.69pt off 2.76pt}]  (236.5,109) .. controls (249.5,161) and (364.5,156) .. (368.5,109) ;
\draw  [color={rgb, 255:red, 0; green, 0; blue, 0 }  ,draw opacity=1 ][line width=1.5]  (323.17,114) .. controls (323.17,95.22) and (354.17,80) .. (392.42,80) .. controls (430.66,80) and (461.67,95.22) .. (461.67,114) .. controls (461.67,132.78) and (430.66,148) .. (392.42,148) .. controls (354.17,148) and (323.17,132.78) .. (323.17,114) -- cycle ;
\draw [color={rgb, 255:red, 0; green, 0; blue, 0 }  ,draw opacity=1 ]   (392.42,80) -- (382.83,80.61) ;
\draw [shift={(380.24,80.77)}, rotate = 356.37] [fill={rgb, 255:red, 0; green, 0; blue, 0 }  ,fill opacity=1 ][line width=0.08]  [draw opacity=0] (12.5,-6.01) -- (0,0) -- (12.5,6.01) -- cycle    ;
\draw [color={rgb, 255:red, 0; green, 0; blue, 0 }  ,draw opacity=1 ]   (305.01,74.94) -- (287.83,75.61) ;
\draw [shift={(289.03,75.56)}, rotate = 357.79] [fill={rgb, 255:red, 0; green, 0; blue, 0 }  ,fill opacity=1 ][line width=0.08]  [draw opacity=0] (12.5,-6.01) -- (0,0) -- (12.5,6.01) -- cycle    ;
\draw [color={rgb, 255:red, 0; green, 0; blue, 0 }  ,draw opacity=1 ]   (204.42,79) -- (187.52,80.35) ;
\draw [shift={(188.59,80.26)}, rotate = 355.44] [fill={rgb, 255:red, 0; green, 0; blue, 0 }  ,fill opacity=1 ][line width=0.08]  [draw opacity=0] (12.5,-6.01) -- (0,0) -- (12.5,6.01) -- cycle    ;

\draw (180,48) node [anchor=north west][inner sep=0.75pt]   [align=left] {\Large $\gamma_3$};
\draw (278,48) node [anchor=north west][inner sep=0.75pt]   [align=left] {\Large $\gamma_2$};
\draw (375,48) node [anchor=north west][inner sep=0.75pt]   [align=left] {\Large $\gamma_1$};
\draw (75,100) node [anchor=north west][inner sep=0.75pt]   [align=left] {\large $\cdots$};

 \end{tikzpicture}
\end{center}
\caption{圈$\gamma_i$的构造。叉号代表黎曼面上的割点，波浪线是割线。\\
Figure 3: The construction of the cycle $\gamma_i$. The cross marks represent branch points on the Riemann surface, and the wavy lines denote branch cuts.}
\label{fig:cycle}
\end{figure}

在此，我们首先假设$m_k$是正实数\footnote{这一设定需要选取合适的多项式$p(z)$。}。我们引入以下函数
\begin{equation}
    \ell_k=\log Y_k(\theta)-m_s\cosh\theta,
\end{equation}
其在${\rm Im}(\theta)<\pi/2$内是解析的。从Y系统，我们可以推导出
\begin{equation}
    \ell_k^{[+]}+\ell_k^{[-]}=\log\Big((1+Y_{k-1})(1+Y_{k+1})\Big).
\end{equation}
借助留数积分，我们进一步得到
\begin{equation}
    K\star (\ell_k^{[+]}+\ell_k^{[-]})=\ell_k,
\end{equation}
其中$K$是核$K(\theta)=1/(2\pi \cosh\theta)$, $\star$表示卷积。由此，我们得到了TBA方程
\begin{equation}
    \log Y_{k}=-m_{k}\cosh\theta+K\star\Big[\log(1+Y_{k-1})+\log(1+Y_{k+1})\Big],\quad k=1,\cdots,n-3,
\end{equation}
这是$A_{n-3}$型的TBA方程\cite{Zamolodchikov:1991et}。

对于一般的多项式$p(z)$，质量$m_k$可以是复数。这时我们需要转动$\theta$，以保证$m_k/\zeta=|m_k|e^{i\phi_k}/\zeta$是正实数。这时的Y函数的渐近行为可以写作
\begin{equation}
    \log Y_k(\theta+i\phi_k)\sim -|m_k|\cosh\theta,\quad \theta\to \pm\infty.
\end{equation}
类似地，我们可以推导出
\begin{equation}
    \log\tilde{Y}_{k}=-|m_{k}|\cosh\theta+K_{s,s-1}\star\log(1+\tilde{Y}_{k-1})+K_{s,s+1}\star\log(1+\tilde{Y}_{k+1}),
\end{equation}
其中$\tilde{Y}_k=Y_k(\theta+i\phi_k)$, $K_{k,j}=K(\theta+i\phi_k-i\phi_j)$。值得强调的是，当$|\phi_{k,k\pm 1}|=\pi/2$时，TBA的核里会出现极点，超越这些极点，我们需要考虑其留数的贡献，并修改TBA方程。这一现象被称为TBA方程的越墙（wall-crossing）效应\cite{Alday:2010vh,Toledo-thesis}\footnote{也参见
J. Toledo的未发表的文章 ``Notes on wall-crossing''。}。为了简化讨论，我们仅考虑$|\phi_k-\phi_{k\pm 1}|$以避免越墙效应的发生。

\section{极小曲面的面积以及TBA的自由能}\label{sec:free-energy}
Wilson圈上的极小曲面的面积由以下作用量描述
\begin{equation}
A=2\int d^2z\partial \vec{Y}\cdot \bar{\partial}\vec{Y}=4\int d^2ze^{2\alpha}.
\end{equation}
从边界条件\eqref{eq:bdYcond}，我们发现$e^{\alpha}$在边界处发散，$A$ 可以分解为有限部分和发散部分，写作
\begin{equation}
A=A_{\rm fin}+A_{\rm div}.
\end{equation}
其中发散部分$A_{\rm div}$为
\begin{equation}
    \begin{aligned}
        A_{{\rm div}}=&A_{{\rm period}}+A_{{\rm cutoff}},\\
        A_{{\rm period}}=&4\int d^{2}z\sqrt{p\bar{p}}-4\int_{\Sigma}d^{2}z\sqrt{p\bar{p}},\\
        A_{{\rm cutoff}}=&4\int_{\Sigma_{0},r<\epsilon}d^{2}z\sqrt{p\bar{p}},
    \end{aligned}
\end{equation}
这里$\Sigma$是一个参考面，$r$为径向方向。$A_{\rm period}$可以通过黎曼双线性公式计算，$A_{{\rm cutoff}}$依存于在大$z$方向的截断（对应靶空间径向方向的截断)。$A_{{\rm cutoff}}$中的有限部分与BDS假设中的散射振幅有类似的构造。对此，我们不做过多解释，详细部分请参考\cite{Alday:2009yn,Alday:2009dv,Alday:2010vh}.

有限部分为$A_{\rm fin}$是$e^{2\alpha}-\sqrt{p\bar{p}}$的积分。利用修改sinh-Gordon方程，可以将$A_{\rm fin}$改写为
\begin{equation}
\begin{aligned}
A_{\rm fin}
&=2\int d^{2}z(e^{2\alpha}+e^{-2\alpha}p\bar{p}+\partial\bar{\partial}\alpha-2\sqrt{p\bar{p}})\\
&=A_{{\rm free}}+\frac{\pi}{2}(n-2),
\end{aligned}
\end{equation}
这里我们利用边界条件计算了$\partial\bar{\partial}\alpha$的积分。
$A_{\rm free}$是
\begin{equation}
    A_{{\rm free}}=2\int d^{2}z(e^{2\alpha}+e^{-2\alpha}p\bar{p}-2\sqrt{p\bar{p}})=\frac{i}{2}\int\lambda dz\wedge\eta,
\end{equation}
其中，我们引入了两个闭形式
\begin{equation}
    \lambda dz=\sqrt{p} dz,\quad \eta=2\sqrt{\bar{p}}\big(\cosh(2\hat{\alpha}-1) \big)d\bar{z}+\frac{1}{\sqrt{p}}(\partial \hat{\alpha})^2 dz.
\end{equation}
借助黎曼双线性公式，我们可以将这个积分写作黎曼面上圈积分的形式
\begin{equation}
    A_{{\rm free}}=-\frac{i}{2}\sum_{a,b}w_{ab}\oint_{\gamma_{a}}\lambda dz\oint_{\gamma_{b}}\eta,
\end{equation}
这里$w_{ab}$是圈的相交数。值得强调的是$\lambda$和$\eta$出现在Y函数的渐近展开中，我们可以从Y函数的小$\zeta$展开提取出所需要的圈积分\footnote{这里$\omega_{ab}\theta^{bc}={\delta_a}^c$。}
\begin{equation}
    -\oint_{\gamma_a}\lambda dz=Z_a,\quad -\frac{1}{2}\oint_{\gamma_b}\eta=\bar{Z}+\sum_{c}\frac{\theta^{bc}}{\pi i}\int\frac{d\zeta^\prime }{\zeta^\prime}\log\Big(1+\hat{Y}_c(\zeta^\prime)\Big),
\end{equation}
其中$Z_a=-\oint_{\gamma_a}\sqrt{p}dz$, $\hat{Y}_{2k}(\zeta)=Y_{2k}(\zeta)$ 以及 $\hat{Y}_{2k+1}(\zeta)=Y_{2k+1}(\zeta e^{-i\pi/2})$。这样，我们就可以将极小面积的非平凡部分用Y函数的展开表示出来
\begin{equation}
    A_{{\rm free}}=-\sum_{a=1}^nZ_{a}\frac{1}{\pi}\int\frac{d\zeta}{\zeta^{2}}\log(1+\hat{Y}_{a}).
\end{equation}
将$\hat{Y}$该写作Y函数的形式，并利用大$\zeta$展开重复同样的计算并取平均，我们得到极小面积的非平凡部分的最终形式\footnote{利用ODE/IM对应，我们可以从线性问题推导出Bethe拟设方程以及相应的非线性积分方程，其自由能同样的可以给出此面积\cite{Fioravanti:2020udo}。}
\begin{equation}
    A_{{\rm free}}=\sum_{a}\int\frac{d\theta}{2\pi}|m_{a}|\cosh\theta\log(1+\tilde{Y}).
\end{equation}
我们发现$A_{{\rm free}}$ 与量子可积系统TBA方程的自由能具有相同的形式。选取多项式$p(z)$，我们可以得到质量$m_a$，进而得到$n-3$个TBA方程。通过（数值）求解TBA方程，我们便得到$\zeta=1,i$时的Y函数，它们将给出相应的交比\eqref{eq:cross-ratio-1},\eqref{eq:cross-ratio-2}，从而决定Wilson圈的具体形状。最终，从TBA方程的自由能，我们便能得到相应的极小曲面的面积。
最后我们强调一下, 本章节的推导仅适用于奇数$n$的情形。当$n$为偶数时，即$4k$胶子散射，在$dw=\sqrt{p}dz\sim (z^{n/2-1}+\cdots+\frac{\tilde{m}}{z})dz$的无穷远将会出现非平凡的monodromy，这将导致一个额外的项$A_{\rm extra}$的出现 \cite{Alday:2009yn} \footnote{十分感谢匿名审稿人指出此问题。}。AdS$_5$中的推广可参见\cite{Yang:2010az,Yang:2010as}。

\section{总结与展望}\label{sec:con}
本文旨在介绍散射振幅/Wilson圈对偶中极小曲面面积的计算，该极小曲面以AdS边界处的类光多边形Wilson圈为边界。鉴于运动方程的非线性以及复杂的边界条件，直接求解运动方程以计算面积十分困难。本文回顾了绕开对运动方程的直接求解，并利用热力学 Bethe 拟设的自由能计算极小曲面面积的方法。由于篇幅所限，本文仅能顾及热力学 Bethe 拟设和散射振幅相关话题的一个侧面。以下是本文相关的后续发展与展望：

{\bf AdS$_5$空间中的极小曲面:}
同样地，我们可以对AdS$_5$的经典弦的运动做Pohlmeyer约化，得到相应的经典偏微分方程组\cite{Alday:2009dv,Burrington:2009bh,Burrington:2011eh}，以及与之等价的特定SU(4)Hitchin系统。从这一Hitchin系统，我们可推导出Y系统以及相应的TBA方程，其自由能同样的描述了极小曲面面积的非平凡部分\cite{Alday:2010vh,Hatsuda:2010vr,Hatsuda:2012pb}。在此框架下，对面积的讨论将更为便捷，使我们能够完整计算出BDS假设在强耦合下的偏离。


{\bf 非平面散射振幅:} 在AdS/CFT对应关系的背景下，非平面散射振幅的研究一直是一项艰巨的任务，其主要困难在于非平面情况下黎曼曲面的高亏格。为了克服这一困难，可将高亏格的黎曼曲面切割成圆盘，以便在这些圆盘上运用平面极限下的成熟技术，随后将它们拼接起来。基于这一思路，Ben-Israel，Tumanov和Sever提出了（切割后的）非平面散射振幅与多条Wilson线的关联函数的对偶性\cite{Ben-Israel:2018ckc}。在强耦合，我们可以从Hitchin系统引入Fock-Goncharov坐标，并推导出TBA方程计算以多条类光周期性Wilson线为边界的极小曲面面积得到切割后的非平面散射振幅\cite{Ouyang:2022sje}\footnote{这种推导TBA方程的做法更具一般性，其也被用到AdS空间中关联函数的计算\cite{Caetano:2012ac}。}。

\begin{figure}[htp]
    \centering
    \includegraphics[width=0.15\linewidth]{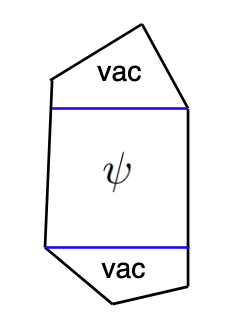}
    \caption{五边形OPE中的Wilson圈分解。\\
    Figure 4: Wilson loop decomposition in pentagon OPE.}
    \label{fig:POPE}
\end{figure}
{\bf 五边形 Operator Product Expansion (OPE):} Wilson圈可以被分割为多个四边形，见图\ref{fig:POPE}。相邻的两个四边形可以组成一个五边形。我们可以将顶部以及底部的四边形理解为Gubser-Klebanov-Polyakov真空态，整个Wilson圈可以被解释为真空态之间的跃迁。这些跃迁与世界面S矩阵紧密相关。利用可积性以及自举法，Alday与Basso等人将提出了一种在任意耦合强度下自举的计算共线（colinear）散射振幅/Wilson圈的方法\cite{Alday:2010ku,Basso:2013vsa}。这种做法被称作五边形OPE。在强耦合极限下，此方法可以重现出AdS$_5$相关的Y系统以及TBA方程\cite{Bonini:2015lfr}。进一步的，五边形OPE被扩展到了形状因子\cite{Sever:2020jjx,Guo:2022qgv}以及非平面散射振幅的情形\cite{Ben-Israel:2018ckc}。

{\bf 对于ODE/IM对应的影响:} 从线性问题构造Y系统以及TBA方程的方法被视为对ODE/IM对应的扩展，受此启发Lukyanov等人发现了修改sinh-Gordon方程和量子sine-Gordon模型的对应\cite{Lukyanov:2010rn}\footnote{目前的研究仅限于$p(z)=z^{2\alpha}-s^{2\alpha}$的情形。}。这种对应被称作线性ODE/IQFT对应或者有质量ODE/IM对应。该对应进一步的被扩展到更高秩\cite{Dorey:2012bx,Ito:2018wgj}以及更一般代数的情形\cite{Ito:2013aea,Ito:2015nla,Ito:2016qzt,Ito:2020htm}。通过对偏微分方程的线性问题取光锥极限（共形极限）\footnote{对应的Hitchin系统的极限见\cite{Gaiotto:2014bza}。}，对应的量子可积系统以及TBA方程将回归到共形场论相关的ODE/IM对应。值得强调的是，受这一系列发展的影响，我们可以从多项式势能的一维量子力学的薛定谔方程出发，推导出描述严格量子周期（quantum period）的TBA方程。利用TBA方程的解析性，我们可以重现量子周期所满足的渐近行为以及非连续性。这一发现极大地促进我们对复现（resurgence）构造的理解，并很快地被扩展到了更一般势能的薛定谔方程\cite{Ito:2018eon,Ito:2019jio,Ito:2024nlt}\footnote{对于Mathieu 方程相关的ODE/IM对应的工作见\cite{Zam-unpub,Grassi:2019coc,Fioravanti:2019vxi}。}以及高阶微分方程的情形\cite{Ito:2021boh,Ito:2021sjo}。更详尽的介绍可参见\cite{Ito:2025pfo}。

\section*{致谢}
HS非常感谢Davide Fioravanti，Katsushi Ito，欧阳昊以及Marco Rossi在相关方向的合作，并感谢Alfredo Bonini，Simone Piscaglia，Yuji Satoh，Amit Sever，杨刚，Dmytro Volin以及王俊凯在相关问题上的讨论。另外，HS非常感谢``Gravity, Field Theories, and Their Relations 2024 -- Integrability: From Mathematics to Experiments''以及``International Workshop on Exact Methods in Quantum Field Theory and String Theory''的支持，本文的撰写在相关学术活动中得到了极大地推进与启发。


\end{CJK*}

\end{document}